# Accurate and Versatile High-Order Modeling of Electromagnetic Scattering on Plasmonic Nanostructures

Hamid T. Chorsi, Stephen D. Gedney*

University of Colorado Denver, Department of Electrical Engineering
Denver, Colorado, USA
hamid.chorsi@ucdenver.edu
stephen.gedney@ucdenver.edu

**Abstract**. This paper presents high-order (HO) electromagnetic modeling of plasmonic nanostructures based on the Locally Corrected Nyström (LCN) method. Advanced nanophotonic and nanoplasmonic structures involve electrically large electromagnetic structures that are very complex in both geometry and material composition. Hence, advanced analysis and design tools are required in order to predict the performance of such structures and optimize the geometrical parameters prior to costly prototype development. In this perspective, the LCN is exploited to solve the electromagnetic scattering of plasmonic nanostructures. The LCN utilizes basis functions of higher orders defined on large geometrical elements, which significantly reduces the number of unknowns for a given problem. Compared to other well-known methods for EM modeling of plasmonic nanostructures, the LCN is computationally efficient, straightforward to implement and provides exponential convergence. A full comparison, in terms of time and complexity between the proposed method and previous works for a few 3D nanoplasmonic structures are presented to validate the accuracy, efficiency and flexibility of the proposed method to simulate the behavior of electromagnetic fields in real nanoplasmonic problems.

## 1. Introduction

In light of recent advances in nanoscience and nanotechnology, metallic nanoparticle structures obtained significant attention due to Surface Plasmon Polariton (SPP). This is the result of coherent coupling of photons to free electron oscillations on the surface of a good metal [1]. This ability leads to several interesting applications, such as breaking the diffraction limit of light, confining optical fields to very small dimensions and directly localized optical sources into the far field. The rise of nanoplasmonic structures have emerged in cutting-edge applications to optical imaging [2, 3], medicine [4], nanoscale light confinement [5], nanolasing and quantum optics [6, 7].

The high cost of fabricating nanophotonic and nanoplasmonic devices causes the design, simulation, and performance verification of the nanophotonic and nanoplasmonic devices to be essential prior to the fabrication of the actual structure. Therefore, numerical methods of nanophotonics have become relevant for solving nanophotonics problems.

Earlier works in this area are primarily based on finite methods. One of the disadvantages of finite methods, such as the finite-difference time-domain (FDTD) [8] or the finite element method (FEM) [9], is that the entire volume of the structure, which is confined to a bounding box with terminating absorbing boundaries, must be discretized. The discretization increases the number of unknowns thus increasing the complexity of the method and simulation time. Although these approaches are becoming increasingly popular, which is due to the simplicity and partially due to the availability of commercial software (e.g., COMSOL, CST and etc.), this method cannot cope with structures spanning several wavelengths in size [10].

A more computationally efficient approach comprises the use of a surface integral equation (SIE) formulation, which has been successfully applied in many RF and microwave applications. Although this method is not yet well received in optics, these methods possess certain important advantages when compared to the aforementioned volumetric approaches. The main advantage is the fact that the approach only discretizes the surface of the object, rather than the entire space containing the material structure. This lowered discretization results in a considerable reduction in the number of unknowns. Furthermore, no absorbing boundary conditions (ABC) or surrounding empty space need to be specifically discretized. In addition, these methods are less sensitive to instabilities produced by instant spatial variations of the permittivity, as is the case in the majority of plasmonic structures [10]. Preceding works in SIE analysis of nanoplasmonic structures have exploited the well-known method of moments (MoM) to solve the perspective integral equations achieved from the SIE formulation [11]. MoM-SIE is an extremely powerful and versatile general computational electromagnetic (EM) tactic and has been applied extensively in electromagnetics for the EM simulations of 2D and 3D objects. However, the MoM is complex and typically requires an expensive double integration [12].

There is always a tradeoff between the solution accuracy and the degrees of freedom representing the discrete surface current densities when simulating large complex geometries. Traditional schemes based on the MOM, classically employ low-order basis functions and testing functions such as the Rao-Wilton-Glisson (RWG), vector basis, that is, a function that leads to fixed error convergence. If increased accuracy is required from the simulation, a significant increase in the computational resources is necessary [13].

In this paper, a high-order analysis method, namely, the Locally Corrected Nyström (LCN) is employed to accurately and efficiently simulate the EM properties of nanoplasmonic structures. The proposed method is applied for the EM scattering analysis of 3D nanoplasmonic structures. A comprehensive error analysis of the proposed method for a canonical structure is presented and the efficiency of the proposed method is demonstrated with comparing results with previous methods. This paper is structured as follows: in section 2, a review of low-order and high-order methods is presented. A brief derivation of the LCN is presented in sections 3. Simulation results are presented in section 4. Section 5 concludes the paper.

## 2. High-Order vs. Low-Order Methods

Obtaining precise solutions to scattering problems of nanoplasmonic structures with controlled accuracy is required in many hands-on applications. These comprise the analysis of nanoantennas, photonic crystals, double-negative materials, metamaterials, nanophotonic semiconductor devices; among others. Traditional methods in this area are predominantly low-order or small-domain techniques in which the structure is modeled by surface geometrical patches (boundary elements) that are electrically very small and the currents (electric and magnetic) are approximated by low-order basis functions such as the rooftop basis [14] or Rao-Wilton-Glisson (RWG) [15]. Moreover, the patches are fractions of the wavelength (usually $\lambda/10$). When the problem is large (electrically), this results in a very large number of unknowns. In addition, these patches are usually plane (flat) triangles or bilinear quadrilaterals which may require dense discretization to accurately model curved structures. Furthermore, the accuracy of solution while using the low-order bases is improved slowly with increases in the number of unknowns ($O(h^2)$ or linear) [16, 17]. Higher-order or large-domain numerical methods elegantly resolve aforementioned problems. According to this method, a structure is approximated by a number of as large as possible geometrical elements (usually curvilinear patches), and the fields (or the current density) are represented by high order (polynomial) basis functions. In contrast to low-order methods, high-order methods are exponentially convergent and can allow one to control the rate of convergence with basis order. Consequently, high-order methods are capable of substantially reducing the solution error at the expense of only a modest increase in computational resources and complexity [13].

## 3. LCN formulation

In this section, the basic Locally Corrected Nyström (LCN) method is summarized. Additional detail on the LCN formulation can be found in [12]. Consider the integral equation used to solve for the surface current density $J(r')$.

$$\int_S K(r,r') J(r') ds' = f^{inc}(r), \quad r \in S \qquad (1)$$

where $S$ is a smooth surface, $r$ is an observation point on the surface $S$, $f^{inc}(r)$ is the known forcing function (excitation field) and $K(r,r')$ is the kernel that may or may not exhibit singular behavior. The surface $S$ is discretized into $N_p$ curvilinear patches that represent the surface outline to high-order. By discretizing (1) over $N_p$ patches we obtain

$$\sum_{p=1}^{N_p} \int_{S_p} K(r,r') J(r') ds' = f^{inc}(r), \quad r \in S \qquad (2)$$

Then a suitable quadrature rule is utilized to approximate the integral over each patch. The observation field is sampled at each of the quadrature points on $S$, leading to

$$\sum_{p=1}^{N_p}\sum_{q=1}^{N_q}\omega_{q_p}K(\mathrm{r}_{q_m},\mathrm{r}'_{q_p})\mathrm{J}(\mathrm{r}_{q_p})=\mathrm{f}^{inc}(\mathrm{r}_{q_m}) \quad (3)$$

where $\mathrm{r}_{q_p}$ and $\mathrm{r}_{q_m}$ are the abscissas on the $p$ th and $m$ th patches, respectively and $\omega_{q_p}$ are weights on the $p$ th patch. In the Nystrom scheme, the current sampled at the quadrature abscissae, $\mathrm{J}(\mathrm{r}_{q_p})$, become the unknowns. Then, sampling the field point at each quadrature abscissa point, leads to a square linear system of equations, with the $q_m$ th row defined by (3). As posed, the Nyström method is limited to regular kernels. For electromagnetic applications most kernels are not regular, and, in fact, are singular. Singularity happens when the observation and source points approach each other in which the kernel in (3) goes to infinity. A novel way to alleviate this problem was presented by Strain [18] in which local corrections are applied to the numerical quadrature rule in near field regions where the distance between the observation and source points is small. Through the local corrections, (3) can be rewritten as

$$\sum_{p\in far}\sum_{q=1}^{N_q}\omega_{q_p}K_{q_m,q_p}\mathrm{J}(\mathrm{r}_{q_p})+\sum_{p\in near}\sum_{q=1}^{N_q}\tilde{\omega}_{q_p}\mathrm{J}(\mathrm{r}_{q_p})=\mathrm{f}^{inc}(\mathrm{r}_{q_m}) \quad (4)$$

Where $K_{q_m,q_p}$ is the discrete kernel and $\tilde{\omega}_{q_p}$ are the weights of the specialized local quadrature rule for the singularity at $\mathrm{r}_{q_m}$.

To solve for (4) one can distribute a set of basis function $\{f_{k_n}(\mathrm{r}),\ k_n=1,...,\mathrm{K}_n\}$ over the $p$ th element, where the subscript $n$ identifies the $K_n$ bases over $S_p$

$$\sum_{q=1}^{N_q}\tilde{\omega}_{q_p(\mathrm{m})}f_{k_n}(\mathrm{r}_{q_p})=\int_{S_p}K(\mathrm{r}_{q_m},\mathrm{r})f_{k_n}(\mathrm{r}')\mathrm{d}s' \quad (5)$$

$$L_p\tilde{\omega}_{q_p(\mathrm{m})}=\tilde{k}_{q_m} \quad (6)$$

where $L_p$ is referred to as the local correction matrix of patch $p$. Using (4) and (6) a square matrix forms which will be used to solve for the $\mathrm{J}(\mathrm{r}')$.

## 4. Simulation Results and Discussion

In this sections, we present numerical examples to demonstrate the versatility and accuracy of the presented formulation in plasmonics. Two surface integral equation (SIE) formulations are considered, namely, the Poggio–Miller–Chang– Harrington–Wu–Tsai (PMCHWT) and the Müller methods [19]. At the outset, the accuracy of the proposed methods are examined by comparing the simulation results with the analytical solution provided by the Mie series for a plasmonic nanosphere. A gold nanosphere with $\varepsilon_r = -11.095 - j1.2603$ and a 200 nm radius that is illuminated from $\theta = 0^o, \phi = 0$ with a vertically polarized plane wave at 467.7644 THz (641 nm) ($\vec{E}^{inc} = \hat{x} e^{jk_0 z}$ V/m). The sphere is discretized with 64, eighth-order quadrilateral cells as shown in Fig. 1(a).

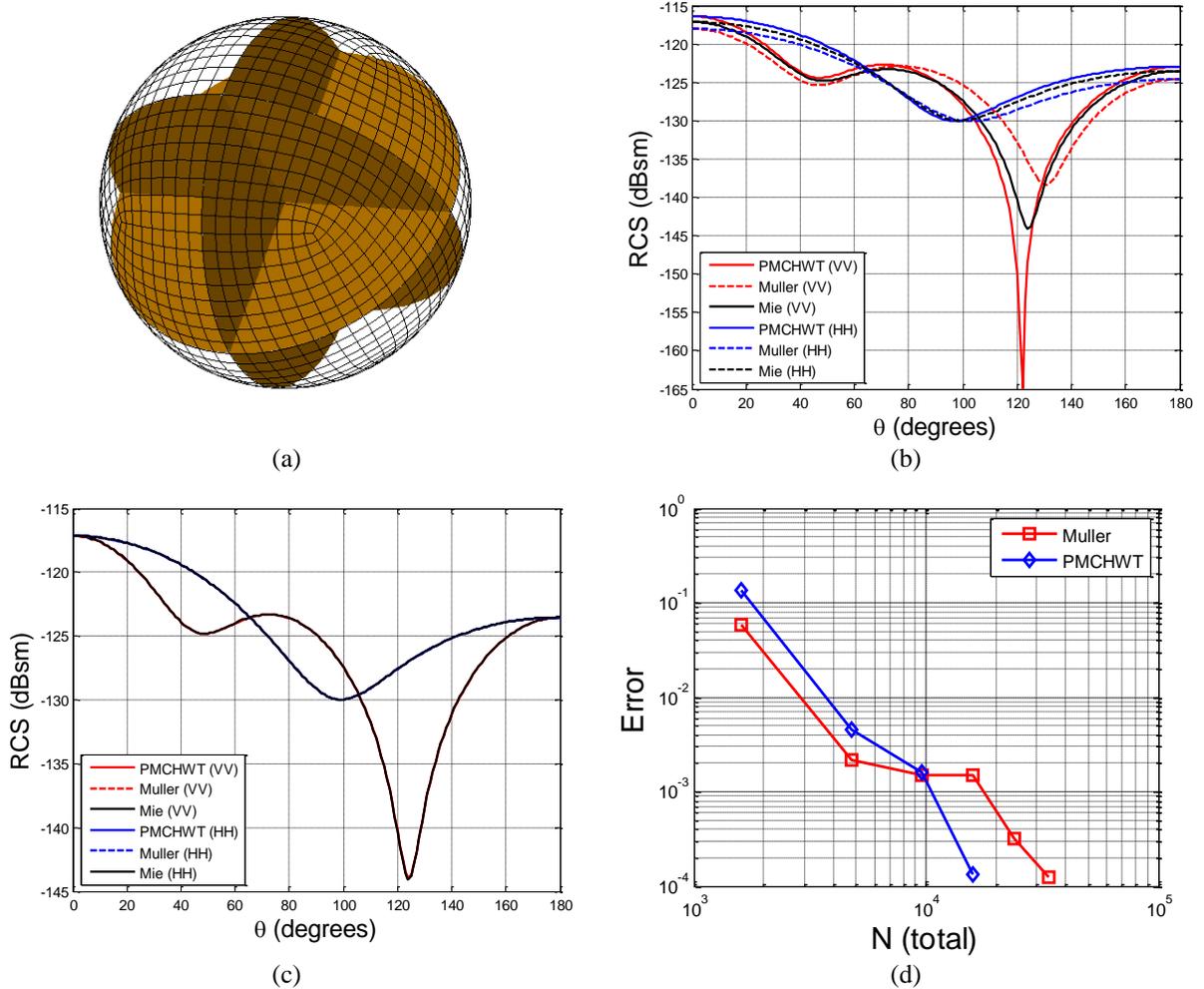

Fig. 1. Bistatic RCS of the gold nanosphere and comparison with Mie series. (a) discretization of nanosphere using curvilinear patches (b) bistatic RCS of nanosphere with p=1 (c) bistatic RCS of nanosphere with p=2 (d) convergence of error for two SIEs (PMCHWT and Muller) as p is increased from 0 to 5.

Fig. 1(b). shows the bistatic RCS of the sphere for the zeroth order discretization of the current densities (p=0). As it can be seen from the Figure, results are not converged yet. Increasing the order of discretization will result in the convergence of the results (curves) as shown in Fig. 1(c). A close agreement can be observed between analytical and simulation results, which confirms the validity and accuracy of the method. Fig. 1(d). shows the plot of the error of the PMCHWT-LCN and Muller-LCN with respect to the order of discretization. It can be seen that the error converges exponentially by increasing the order of mixed-order basis function.

We now prove that the results of the analysis conducted so far can be extended to the scattering from nanoparticles featuring complex shapes. For this purpose consider the problem of the scattering from a bowtie nanoantenna. The bowtie structure is extremely attractive for nanoantenna applications because of its triangular geometry, which leads to the "lightning-rod" effects at the gap apexes [20].

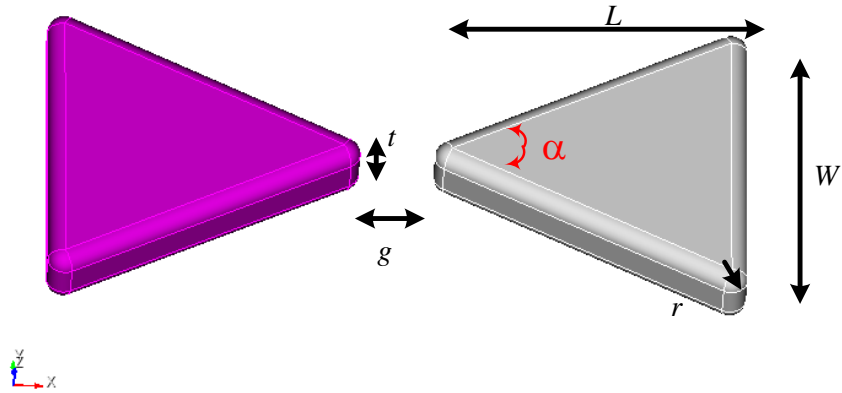

Fig. 2. Gold, nano-bowtie antenna. $L$ = length, $W$ = width, $t$ = thickness, $g$ = gap distance, $r$ = radius of curvature of corners, $\alpha$ = interior angle.

Figure 2 illustrates the structure and geometric parameters of the bowtie nanoantenna. For sake of illustration, a bowtie structure with the dimensions L = 50 nm, W = 50 nm, t = 10 nm, g =11.98 nm, r = 5 nm, α= 53.13°, is considered. Figure 3. Shows the bi-static RCS of the presented bowtie nanoantenna when it is illuminated with vertically and horizontally polarized plane waves with $\left(\theta^{inc},\phi^{inc}\right)=(0°,0°)$ at a frequency of f = 467.7644 THz $(\lambda=641\text{ nm})$.

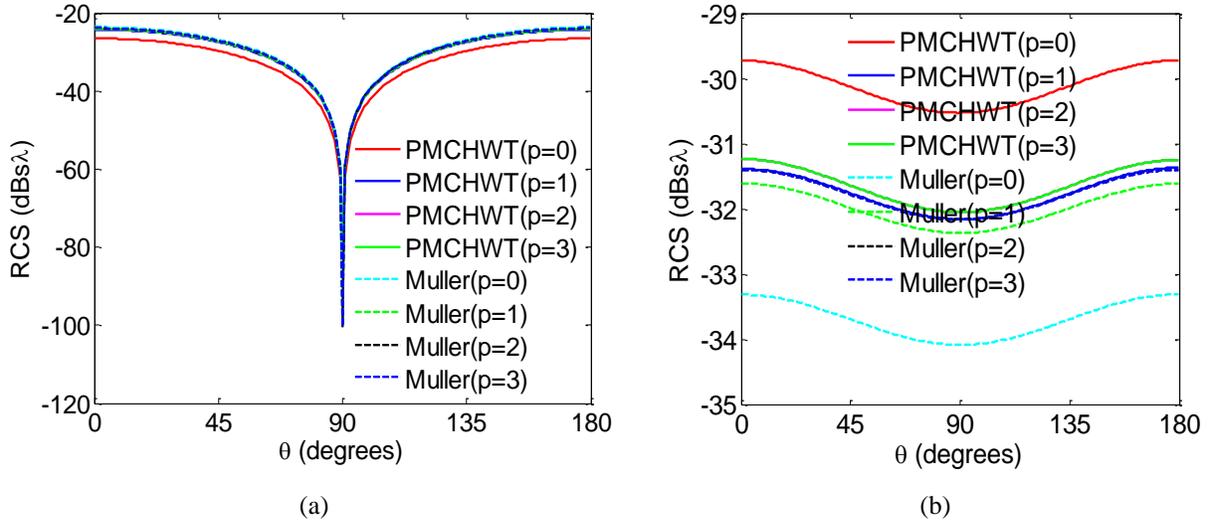

Fig. 3. Bistatic RCS of the bowtie in the $\phi = 0°$ plane with f = 467.7644 THz, $\left(\theta^{inc}, \phi^{inc}\right) = (0°, 0°)$, PMCWT and Müller formulation, for a discretization with 226 quadratic quadrilateral cells, LCN solution. (a) $\sigma_{V-V}$, (b) $\sigma_{H-H}$

A complete comparison between the proposed method and low-order counterparts (MoM-via-RWG) is presented below to show the efficiency of the proposed method. From Figure 4 one can see the exponential convergence of the LCN compared to the MoM for both SIE formulations.

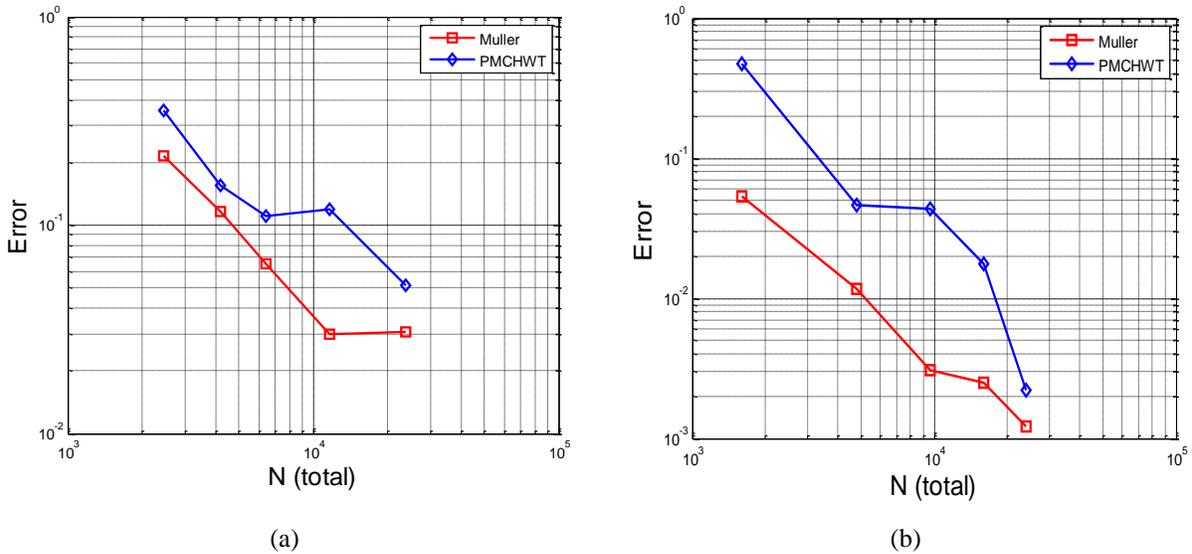

Fig. 4. Relative Error in RCS (MoM), Relative Error in RCS (200 cell, LCN)

Figure 5. Presents the surface magnetic current density for PMCHWT formulation with p=2 due to normally incident vertically polarized plane wave. Form Fig. 5(b). One can notice that the results

from high-order discretization (using high-order patches) is very close to the result using low-order (small patches). Therefore, using high-order discretization, equivalent results are obtained compared to low-order discretization, this decreases the number of unknowns and thus reduces the simulation time.

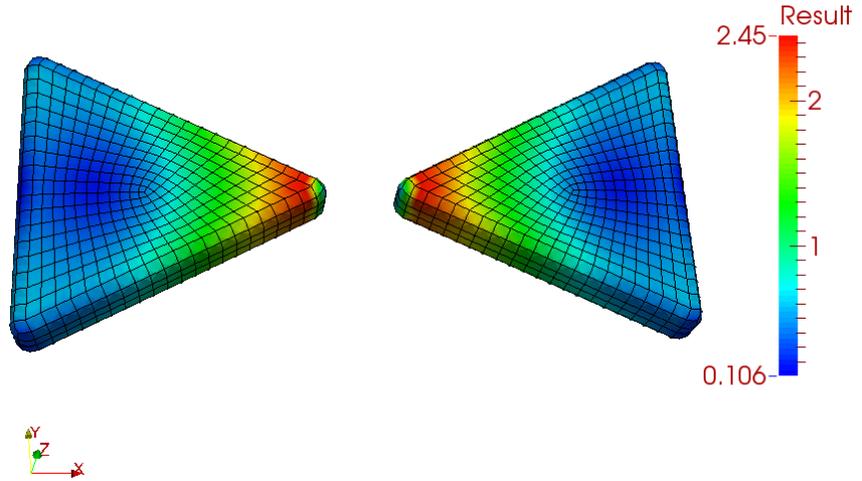

(a)

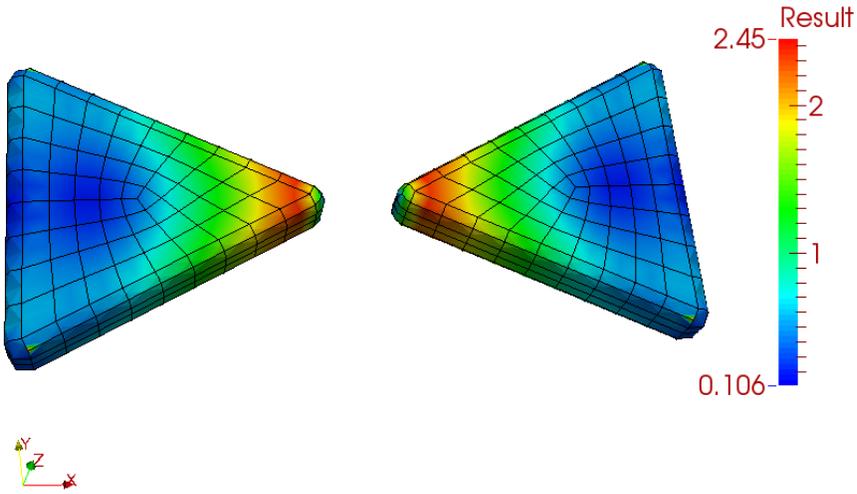

(b)

Fig. 5. Surface magnetic current density $\vec{M} = \vec{E} \times \hat{n}$, 467.7644 THz ($\lambda = 641$ nm), $\vec{E}_\theta^{inc}$ with $\left(\theta^{inc}, \phi^{inc}\right) = \left(0°, 0°\right)$, PMCWT formulation, (a) 1768, and (b) 540 quadratic quadrilateral cells, LCN solution with $p = 2$.

Next, the effect of varying geometrical parameters (gap size, bow angle and length) on the resonance frequency of the bowtie nanoantenna is studied using the LCN simulations based on the

Muller method. Figures 6, 7, and 8 show the electric field enhancement in the gap for different geometrical parameters.

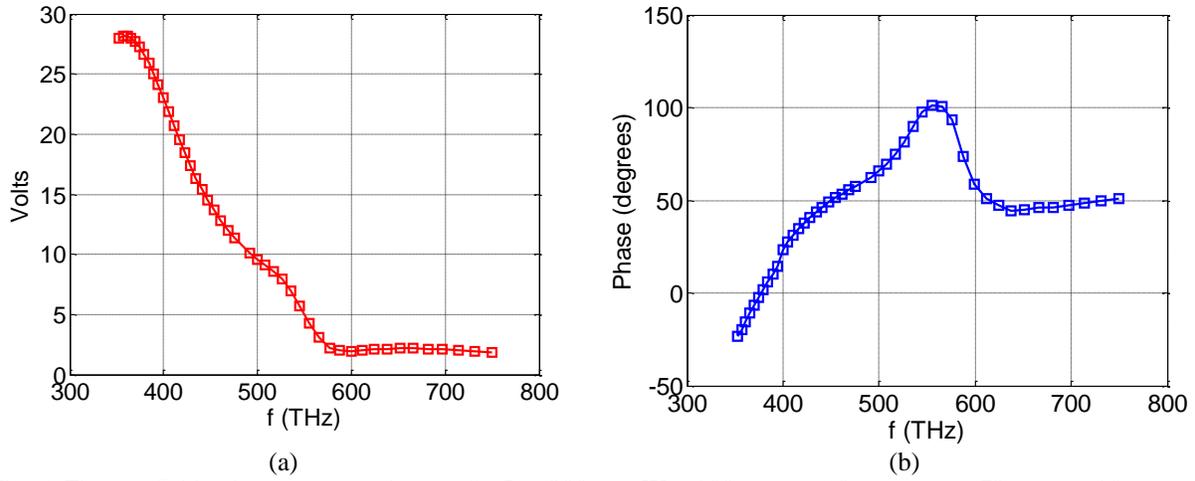

Fig. 6. Electric field enhancement in the gap for L = 200 nm, W = 100 nm, t = 40 nm, g = 44.72 nm, r = 10 nm, α = 53.13° (a) magnitude. (b) phase.

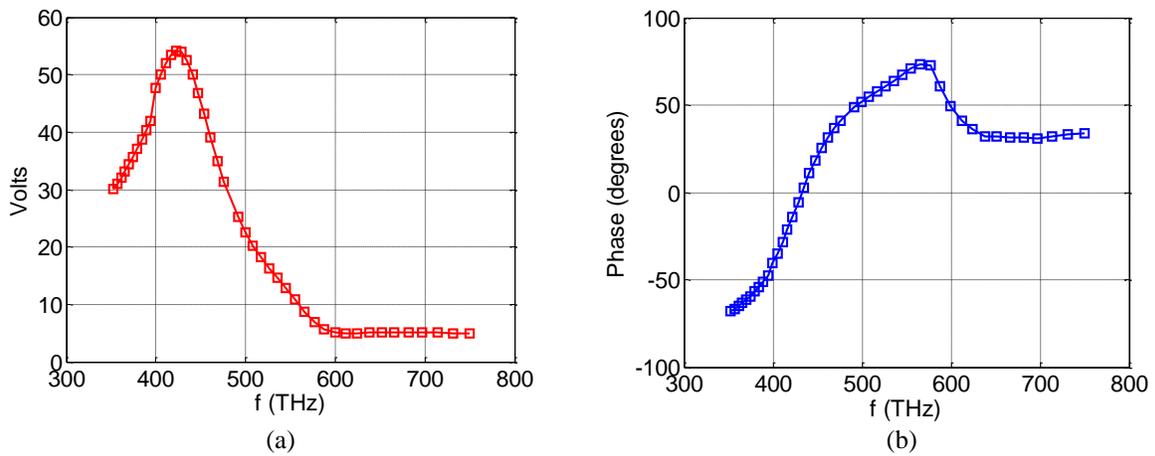

Fig. 7. Electric field enhancement in the gap for L = 150 nm, W = 150 nm, t = 40 nm, g = 34.72 nm, r = 10 nm, α = 53.13° (a) magnitude. (b) phase.

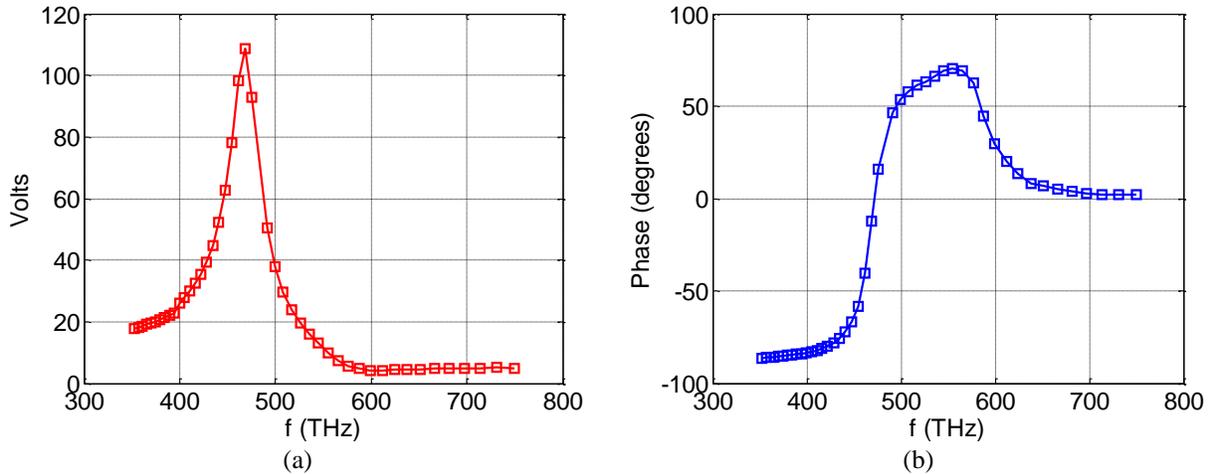

Fig. 8. Electric field enhancement in the gap for L = 50 nm, W = 50 nm, t = 10 nm, g =11.98 nm, r = 5 nm, α= 53.13˚ (a) magnitude. (b) phase.

What is observed from above figures is that the geometric dimensions of bowtie has a great impact on the resonant properties. In this instance, reducing *L, W*, and *t*, while properly adjusting *g*, increases the first frequency as well as the quality factor.

Next consider a more complex structure, a Yagi–Uda nanoantenna which is embedded in glass ( $\varepsilon_r = 2.25$ ). Yagi-Uda nanoantennas have many applications in photovoltaics and optical imaging [21]. Figure 9 shows the geometry of the Yagi–Uda nanoantenna.

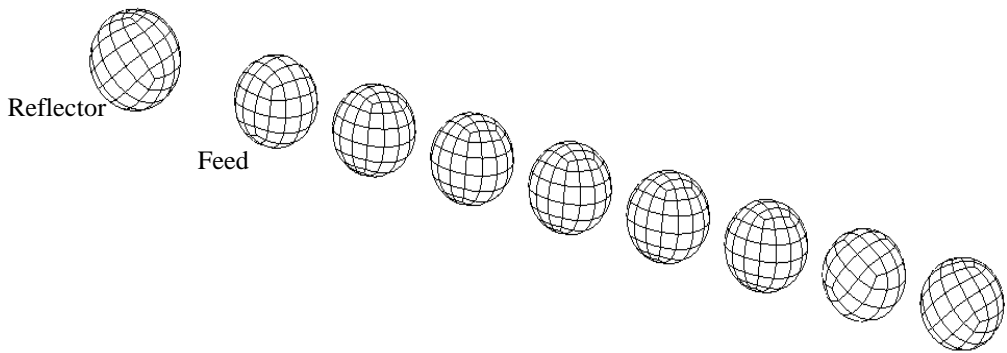

Fig. 9. Surface mesh of the Yagi–Uda antenna (quadrilateral curvilinear patches) embedded in glass. The antenna is made of spherical silver nanoparticles

The antenna consists of one reflector sphere with radius of 60nm and an array of eight director spheres with radius of 55nm and spaced 150 nm. The distance between the reflector sphere and the first director sphere is 175 nm, which gives maximum directivity. A dipole excitation is placed between the reflector and the feed spheres, at a distance of 40nm from the reflector in +z direction.

The antenna is made of silver ($\varepsilon_r = -20.09 - j0.45$ at an operating wavelength of 650 nm). Figures 10 and 11 show the surface electric and magnetic current densities for the Yagi–Uda nanoantenna, respectively.

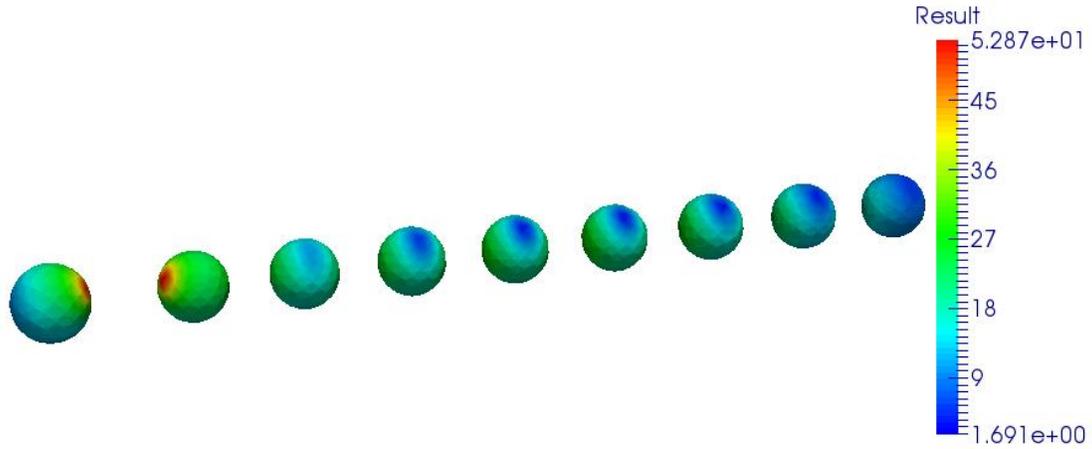

Fig. 10. Surface electric current density ($\vec{J} = \hat{n} \times \vec{H}$) on the surface of the Yagi-Uda nanoantenna due to dipole excitation using the LCN-Muller, cell size=20 nm, p=2

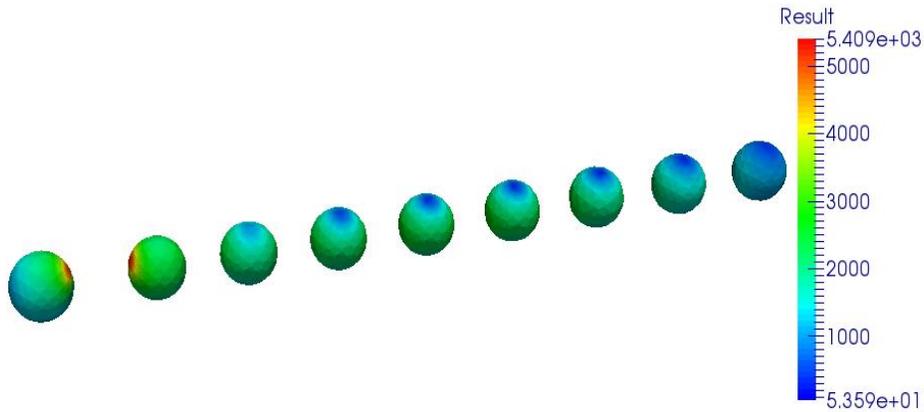

Fig. 11. Surface magnetic current density ($\vec{M} = \vec{E} \times \hat{n}$) on the surface of the Yagi-Uda nanoantenna due to dipole excitation using the LCN-Muller, cell size=20 nm, p=2

Results in Figures 10 and 11 are obtained by using high-order LCN. Results show that the proposed method are capable of accurately modeling the surface electric and magnetic current densities on Yagi-Uda nanostructure.

Table is provided in order to show the time complexity of the LCN method with different order (p) mixed-order basis functions and mesh sizes. As mentioned in [22], if higher order basis functions for currents/fields are used with low-order discretization, then higher order basis functions actually reduce to low-order functions. For example for LCN-Muller formulation even though we are using p=4 but since we are using small domain discretization (41) the simulation time is more that the case in which we use p=2 with high-order discretization (20).

| Method | SIE-Method | Mesh size | $p$ | N | CPU-sec (Fill Time) (seconds) | CPU-sec (Factor Time) (seconds) |
|---|---|---|---|---|---|---|
| LCN | Müller | 10 | 0 | 30104 | 648.005 | 502.766 |
| LCN | Müller | 13 | 0 | 16432 | 212.02 | 86.5524 |
| LCN | Müller | 13 | 1 | 49296 | 1490.92 | 11529 |
| LCN | Müller | 20 | 0 | 7488 | 62.0864 | 9.77977 |
| LCN | Müller | 20 | 1 | 22464 | 197.546 | 212.237 |
| LCN | Müller | 20 | 2 | 44928 | 606.696 | 3326.84 |
| LCN | Müller | 41 | 0 | 960 | 4.7495 | 0.280053 |
| LCN | Müller | 41 | 1 | 2880 | 8.39326 | 0.734545 |
| LCN | Müller | 41 | 2 | 5760 | 17.3173 | 4.13662 |
| LCN | Müller | 41 | 3 | 9600 | 31.6754 | 18.4449 |
| LCN | Müller | 41 | 4 | 14400 | 62.7672 | 58.3088 |
| LCN | PMCHWT | 10 | 0 | 30104 | 1203.99 | 263.971 |
| LCN | PMCHWT | 13 | 0 | 16432 | 412.279 | 64.3253 |
| LCN | PMCHWT | 13 | 1 | 49296 | 1605.61 | 2016.61 |
| LCN | PMCHWT | 20 | 0 | 7488 | 125.302 | 6.55071 |
| LCN | PMCHWT | 20 | 1 | 22464 | 362.433 | 163.799 |
| LCN | PMCHWT | 20 | 2 | 44928 | 987.528 | 1598.33 |
| LCN | PMCHWT | 41 | 0 | 960 | 9.43707 | 0.0661393 |
| LCN | PMCHWT | 41 | 1 | 2880 | 16.0838 | 0.825621 |
| LCN | PMCHWT | 41 | 2 | 5760 | 30.2666 | 3.89454 |
| LCN | PMCHWT | 41 | 3 | 9600 | 49.9672 | 11.919 |
| LCN | PMCHWT | 41 | 4 | 14400 | 91.9565 | 42.9499 |
| MoM | Müller | 10 | 0 | 23580 | 3925.37 | 238.998 |
| MoM | Müller | 13 | 0 | 13092 | 1220.35 | 42.2352 |
| MoM | Müller | 20 | 0 | 5688 | 236.696 | 3.60913 |
| MoM | Müller | 41 | 0 | 1986 | 33.2599 | 0.278411 |
| MoM | Müller | 61 | 0 | 972 | 10.0881 | 0.0403196 |
| MoM | PMCHWT | 10 | 0 | 23580 | 3845.08 | 235.566 |
| MoM | PMCHWT | 13 | 0 | 13092 | 1203.62 | 40.5483 |
| MoM | PMCHWT | 20 | 0 | 5688 | 235.045 | 3.59488 |
| MoM | PMCHWT | 41 | 0 | 1986 | 32.7861 | 0.306204 |

Table 1. Comparison between the LCN and the MoM in terms of time and complexity

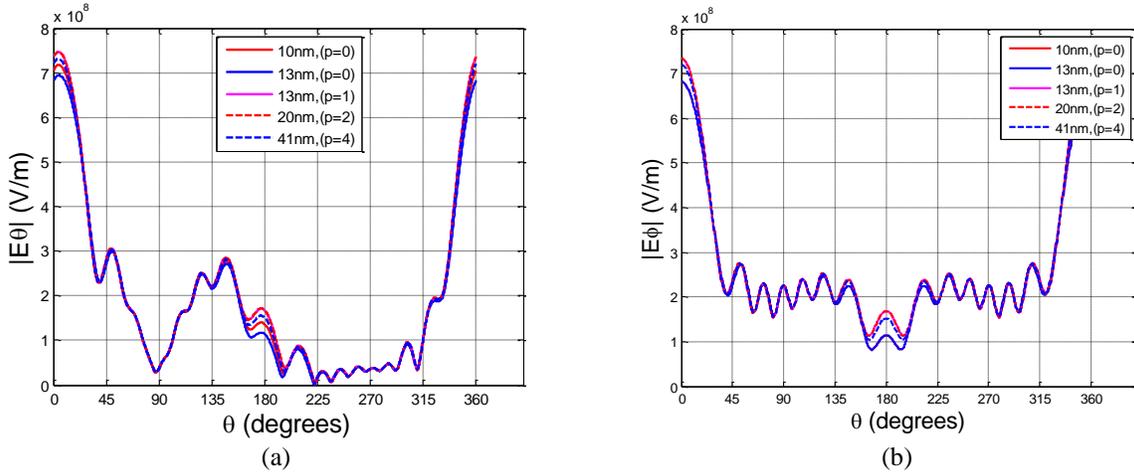

Fig. 12. LCN results, (a) Centered Dipole, E-plane (phi = 0), Muller, (b) Centered Dipole, H-plane (theta = 90), Muller

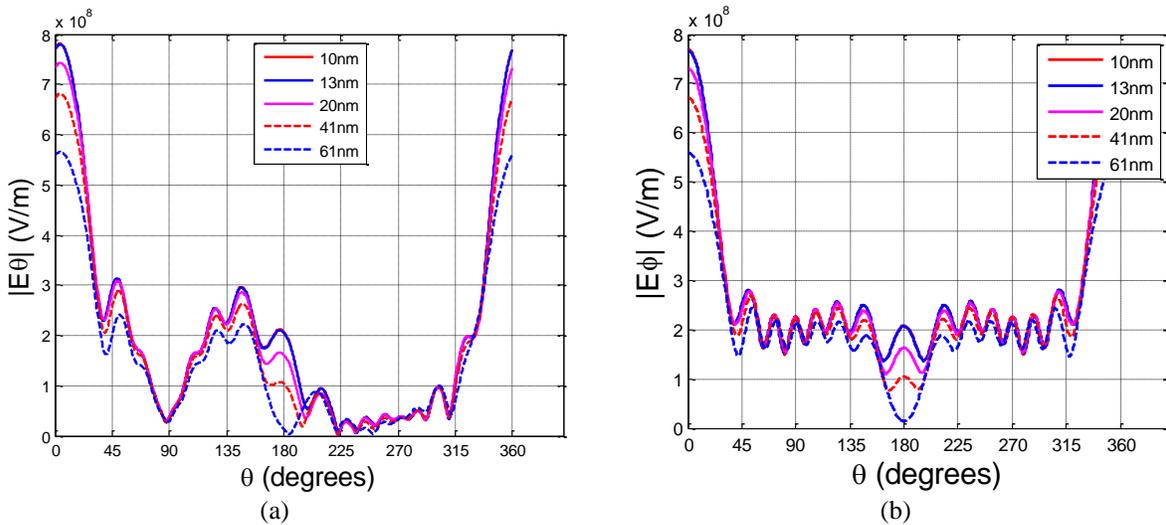

Fig. 13. MoM results, (a) Centered Dipole, E-plane (phi = 0), Muller, (b) Centered Dipole, H-plane (theta = 90), Muller

From Table 1 one can conclude that the PMCHWT formulation is faster that Muller formulation, this is a valid conclusion but as we presented for the gold sphere (Fig. 1(d)), the error convergence of the Muller formulation is faster than the PMCHWT formulation. Table1 also presents a comparison between the LCN and the MoM. Considering CPU times for computing (fill) and solving (factor) the system matrix, one can see the efficiency of the LCN method in contrast to the MoM method.

Observing the LCN results in Figure 12, it is very interesting that everything has converged away from 0 and 180 very quickly. On the other hand, the MoM is monotonically converging in Figure 13. This example shows the promise of the LCN, but also a need for some further work.

## 5. Conclusion

In this paper, a high-order analysis method, namely the Locally Corrected Nyström (LCN), has been applied to efficiently solve the time-harmonic electromagnetic scattering by three-dimensional plasmonic nanostructures. Using a rigorous study of accuracy and time it has been shown that the LCN method is an accurate and efficient method for simulating nanophotonic and nanoplasmonic structures. Through the study of several examples, it has been found that the proposed method is versatile and can be used to model electromagnetic fields on various nanoplasmonic structures. The proposed method can be used to design and optimize future nanoplasmonic applications such as plasmon-enhanced light emitters, biosensors, photodetectors, nanoantennas, and novel probes for optical near-field spectroscopy.